\documentclass[english]{IEEEtran}
\usepackage[T1]{fontenc}
\usepackage{babel}
\usepackage{url}
\usepackage{amsthm}
\usepackage{amsmath}
\usepackage{amssymb}
\usepackage{color}
\usepackage{graphicx}
\usepackage{enumerate}
\usepackage[unicode=true]{hyperref}
\usepackage{comment}
\usepackage{todonotes}
\usepackage[normalem]{ulem}
\usepackage{upgreek}
\makeatletter

\providecommand{\tabularnewline}{\\}

\theoremstyle{plain}
\newtheorem{thm}{\protect\theoremname}
\theoremstyle{plain}
\newtheorem{lem}[thm]{\protect\lemmaname}
\theoremstyle{plain}
\newtheorem{corollary}[thm]{\protect\corollaryname}
\theoremstyle{plain}
\newtheorem{prop}[thm]{\protect\propositionname}
\theoremstyle{definition}
\newtheorem{defn}[thm]{\protect\definitionname}

\theoremstyle{plain}

\usepackage[ruled,linesnumbered]{algorithm2e}
\usepackage{bbm,xspace}

\makeatother

\providecommand{\definitionname}{Definition}
\providecommand{\lemmaname}{Lemma}
\providecommand{\propositionname}{Proposition}
\providecommand{\theoremname}{Theorem}
\providecommand{\corollaryname}{Corollary}

\newcommand{\AAra}[1]{\textcolor{blue}{\bf #1}}

\newcommand{\varalgoname}{SDCP\xspace}    
\newcommand{\varalgonamelong}{Stochastic Dynamic Cache Partitioning }
\newcommand{\varstorage}{K}
\newcommand{\varnCPs}{P}
\newcommand{\varallocation}[1]{\theta_{#1}}
\newcommand{\varallocvec}{\boldsymbol{\uptheta}}
\newcommand{\varallocdiscr}{\hat{\varallocvec}}
\newcommand{\varfeasibleallocs}{\Theta}
\newcommand{\varCPindex}{p}
\newcommand{\varmissintensity}{L}
\newcommand{\varmissintensitylinvec}{\bar{L}}
\newcommand{\varmissintensityvec}{\vec{\mathbf{L}}}
\newcommand{\vartimeslotlength}{T}
\newcommand{\vartimeslot}{k}
\newcommand{\varallocationset}{\mathcal{C}}
\newcommand{\varcenterpt}{\gamma}
\newcommand{\varcenterptvec}{\boldsymbol{\Gamma} }
\newcommand{\varperturbvec}{\mathbf{D}}
\newcommand{\varperturb}{D}
\newcommand{\varmeasuredmiss}{y}
\newcommand{\varmeasuredmissvec}{\vec{\mathbf{\varmeasuredmiss}}}
\newcommand{\vardmeasuredmiss}{\delta\varmeasuredmiss}
\newcommand{\vardmeasuredmissvec}{\delta\varmeasuredmissvec}
\newcommand{\vargradientvec}{\mathbf{\hat{g}}}
\newcommand{\varstepsize}{a}
\newcommand{\varprojection}{\boldsymbol{\upvarphi}}
\newcommand{\varsubgradient}{\bar{\mathbf{g}}}
\newcommand{\varfwddiff}{\Delta}
\newcommand{\varnoisevec}{\boldsymbol{\upvarepsilon}}
\newcommand{\varnonnegspace}{\mathbb{R}^{\varnCPs}_{\geq 0}}
\newcommand{\vararrivals}{A}
\newcommand\varuniformbound{c}

\newcommand\varmissratio{m}
\newcommand\defeq{\triangleq}
\newcommand\varconditionalexp{\epsilon}
\newcommand\expectation{\mathbb{E}}

\newcommand{\lthm}{Theor.~}
\newcommand{\lsec}{Sec.~}
\newcommand{\ldef}{Def.~}
\newcommand{\lalg}{Alg.~}
\newcommand{\lcor}{Cor.~}
\newcommand{\lfig}{Fig.~}

\newcommand{\rdef}[1]{\ldef\ref{def:#1}}
\newcommand{\rlem}[1]{Lemma~\ref{lem:#1}}
\newcommand{\rthm}[1]{\lthm\ref{thm:#1}}
\newcommand{\ralg}[1]{\lalg\ref{alg:#1}}
\newcommand{\rfig}[1]{\lfig\ref{fig:#1}}
\newcommand{\rcor}{\lcor}
\newcommand{\varsapint}[1]{\mbox{SAP}_{\mbox{int}}(#1)}
\newcommand{\varsapcont}[1]{\mbox{SAP}_{\mbox{cont}}(#1)}

\newcommand{\emailcommand}[1]{#1}

\newcommand{\censure}[1]{}  

\newcommand{\atstep}[1]{^{(#1)}}

\newcommand\mycommfont[1]{\footnotesize\ttfamily{#1}}
\SetCommentSty{mycommfont}

\begin{document}
\bstctlcite{IEEEexample:BSTcontrol}

\title{\varalgonamelong\\ for Encrypted Content Delivery}
\author{
\IEEEauthorblockN{Andrea Araldo\IEEEauthorrefmark{1}\IEEEauthorrefmark{3}, Gy\"{o}rgy D\'{a}n\IEEEauthorrefmark{2}, Dario Rossi\IEEEauthorrefmark{3}}\\
\IEEEauthorblockA{\IEEEauthorrefmark{1} LRI, Universit\'e ParisSud, Orsay, France -- \emailcommand{araldo@lri.fr}}\\
\IEEEauthorblockA{\IEEEauthorrefmark{2} School of Electrical Engineering, KTH Royal Institute of Technology, Stockholm, Sweden -- \emailcommand{gyuri@ee.kth.se}}\\
\IEEEauthorblockA{\IEEEauthorrefmark{3} T\'el\'ecom ParisTech, Paris, France -- \emailcommand{dario.rossi@telecom-paristech.fr}}
}

\maketitle

\begin{abstract}
  In-network caching is an appealing solution to cope with the increasing bandwidth demand 
  of video, audio and data transfer over the Internet. Nonetheless, an increasing
  share of content delivery services adopt encryption through HTTPS, which is not 
  compatible with traditional ISP-managed approaches like transparent and proxy caching. 
  This raises the need for solutions involving both Internet Service Providers (ISP) and Content Providers (CP): by design, the solution should preserve business-critical CP information (e.g., content popularity, user preferences) on the one hand, while allowing for a deeper integration of caches in the ISP architecture (e.g., in 5G femto-cells) on the other hand.

  In this paper we address this issue by considering a content-oblivious ISP-operated cache.
  The ISP allocates the cache storage to various content providers so as to maximize the bandwidth
  savings provided by the cache: the main novelty lies in the fact that, to protect business-critical information, ISPs only need to measure the aggregated miss rates of the individual CPs and do not need to be aware of the objects that are requested, as in classic caching.
  We propose a cache allocation algorithm based on a perturbed stochastic subgradient method, and prove that the algorithm converges close to the allocation that maximizes the overall cache hit rate.  We use extensive simulations to validate the algorithm and to assess its convergence rate under stationary and non-stationary content popularity. Our results (i) testify the feasibility of  content-oblivious caches and (ii) show that  the proposed algorithm can achieve within 10\% from the global optimum in our evaluation.  

\end{abstract}

\section{Introduction}

It is widely known that content delivery over the Internet represents a sizeable and increasing fraction of the overall traffic demand. Furthermore most of the content, including video, is carried over HTTP connections: this evolution of the last decade was not among those forecasted for the IP hourglass model evolution~\cite{deeringIETF51}, and is rather a choice of practical convenience. 
This evolution has a tremendous practical relevance, to the point that  HTTP was overtly recognized
and proposed~\cite{popa2010http} as the new de facto ``thin waist'' of the TCP/IP protocol family. 

In very recent times, we are on the verge of yet another shift of the thin waist: we indeed observe that  the fraction of traffic delivered through HTTPS has already passed 50\%~\cite{Naylor2014}, and it is expected to increase, as the IETF Internet Architecture Board (IAB)  recommends ``protocol designers, developers, and operators to make encryption the norm for Internet traffic''~\cite{IAB2014}.
Besides the IAB recommendation, Content Providers (CP) are already heavily relying on encryption to both protect the privacy of their users, as well as sensitive information (related to user preferences) of their own business.

This evolution toward an all-encrypted Internet creates a tussle between security and efficiency.
Today's Internet heavily relies on middleboxes such as NATs (to combat the scarcity of IPv4 addresses) and transparent or proxy caches~\cite{Barish2000} (to relieve traffic load). However, some of these middleboxes will simply fail to operate in tomorrow's Internet with end-to-end encryption: for example, end-to-end encryption renders caching useless, since multiple transfers of the same object generate different streams if the same object is encrypted with different keys.
At times where the design of the new 5G architecture strives to reduce latency, increase the available bandwidth and better handle mobility, this tradeoff is especially unfortunate, as distributed caches represent a natural way to reduce latency, reduce bandwidth usage and to cope with mobility avoiding long detours to anchor points. 

This architectural evolution calls for a redesign of the current operations involving both Internet Service Providers (ISP) and Content Providers (CP): by design, the solutions should preserve business-critical CP information (e.g., content popularity, user preferences) on the one hand, while allowing for a deeper integration of caches in the ISP architecture (e.g., in 5G femto-cells) on the other hand.

In this paper we address this issue by proposing a content-oblivious algorithm that manages the storage space of an ISP cache that delivers \emph{encrypted content}: the algorithm \emph{dynamically partitions} the cache storage among various CPs so as to maximize the cache hit rate (hence, the bandwidth savings). The most important feature of the algorithm is that in order to protect business-critical information of the CPs, the ISP only needs to measure the \emph{aggregated miss rates} of the individual CPs. We prove that relying on the aggregated miss rates only, our algorithm converges close to the optimal allocation that maximizes the overall cache hit rate, and provide a bound on the gap compared to the optimal allocation. Extensive simulations under realistic scenarios show the feasibility and good performance of the proposed algorithm.

The rest of the paper is organized as follows. \lsec\ref{sec:related} reviews related work. \lsec\ref{sec:model} describes the system model and presents our cache partitioning algorithm. \lsec\ref{sec:analysis} provides analytic results, and \lsec\ref{sec:numerical} evaluates the performance of the algorithm.
\lsec\ref{sec:conclusion} concludes the work.



\section{Related Work}\label{sec:related}

The problem of cache partitioning has first been considered 
in the context of partitioning CPU cache among competing processes~\cite{Wolf1992}.
Clearly, the cache workload created by CPU jobs differs very much from the characteristics of
traffic requests from a user aggregate, so that our technique significantly differs in their design choices: i.e., there is no notion of privacy within the CPU that would force the scheduler to be agnostic of the instruction patterns, unlike in our case.

Related to our work are recent works in the area of in-network caching, which is an established technique that continues to receive significant interest in recent years~\cite{Zhang2013a}.  
They generally consider the cache as a whole, but for 
a few  exceptions~\cite{Carofiglio2013b,Araldo2016} that partition a cache among different applications and video quality levels, respectively, and assume that the served content is observable by the ISP.
Our work differs from these works, as we consider a cache partitioned among CPs, and the ISP can observe the aggregate miss rates of the CPs only.
A recent work~\cite{Hoteit} proposes to share a cache managed by an ISP among different CPs in order to achieve fairness. Partitioning of the storage is done using a pricing scheme based on the value that each CP gives to cache space. Unlike~\cite{Hoteit}, our proposed algorithm aims to maximize the cache hit rate, and does not involve payments, which may make its adoption less controversial considering the disputes among ISPs and CPs in recent years~\cite{Gravey2013}. To the best of our knowledge, our work is the first to propose an effective way to maximize the efficiency of an ISP-managed cache, by partitioning it among multiple CPs, without being aware of the content that the CPs serve and without involving payments.

Closer in  scope to ours are worth mentioning  a set of recent works that target  ISP/CDN cooperation~\cite{DiPalantino2009,Jiang2009,Cho2011,Frank2013}. These  not only show that  ISPs have strong incentive in investing in caching to reduce the traffic on their critical paths, but also show that the other Internet actors, i.e. CPs and users, would benefit from ISP in-network caching.  The game theoretical study in~\cite{DiPalantino2009} shows that caches are inefficient when operated by CPs, since CP content placement and ISP traffic engineering are often not compatible. Solutions are proposed in~\cite{Jiang2009,Frank2013}, which however require ISPs to share with the CP confidential information, such as topology, routing policies or link states, and as such are arguably highly impractical. Conversely, \cite{Cho2011} fosters an ISP-operated cache system, but requires the ISP to be able to observe every object requested by the users, which is arguably equally impractical since CPs purposely hide this confidential information via HTTPS. In contrast with these previous works, our solution does not yield to any leaking of business critical information. Furthermore, our solution is not limited to a  single CP, unlike~\cite{Jiang2009}.


Finally, from a technical viewpoint, our work is aligned with recent industry efforts in the Open Caching Working Group (OCWG)~\cite{OpenCache2015}. The mission of the OCWG is to develop standards, policies and best practices for a new layer of content caching within ISP networks, which can coexist with HTTPS and provide shared access to storage for many CPs. Our work fits the OCWG requirements and as such is, we believe, of high practical relevance.

\section{\varalgonamelong}\label{sec:model}
\label{system_model}

\subsection{System Model and Problem Formulation}
We consider a cache with a storage size of $\varstorage$ slots (e.g., in units of MB) maintained by an operator
and shared by $\varnCPs$ content providers (CPs). The operator is not aware of what content the individual slots store and only decides how to partition the slots among CPs.

We denote by $\varallocation{\varCPindex}\in \mathbb{Z}_{\geq0}$ the number
of cache slots allocated to CP $\varCPindex$, which it can use for caching its most popular contents. We define the set of feasible cache allocation vectors 
	\begin{equation}
		\label{eq:theta}
		\varfeasibleallocs \defeq
		\{\varallocvec\in\mathbb{Z}_{\geq0}^{\varnCPs}\vert \sum_{\varCPindex=1}^{\varnCPs}{\varallocation{\varCPindex}}\leq\varstorage \}\subset\mathbb{Z}_{\geq0}^{\varnCPs}.
	\end{equation}

We consider that the arrival of requests for content can be modeled by a stationary process,
and the number of arrivals over a time interval of length $\vartimeslotlength$ can be \emph{bounded} by some positive constant $\vararrivals(\vartimeslotlength)$.
This assumption is reasonable as requests are generated by a finite customer population, and each customer can generate requests at a bounded rate in practice. Upon reception of a request for a content of the CPs that share the cache, the request can either generate a cache hit (for content stored in the CP partition at time of the request) or a cache miss (otherwise). 
 Formally, we denote the \emph{expected} cache miss rate (i.e., expected number of misses per time unit) of CP $\varCPindex$ when allocated $\varallocation{\varCPindex}$ slots of storage by $\varmissintensity_{\varCPindex}(\varallocation{\varCPindex})$. 
We make the reasonable assumption that $\varmissintensity_{\varCPindex}$ is decreasing and strictly convex on $[0\ldots K]$. This assumption corresponds to that having more storage decreases the miss intensity (in expectation) with a decreasing marginal gain, and each CP would in principle have enough content to fill the entire storage.
For convenience we define the cache miss intensity vector
	\begin{equation}
		\label{eq:varmissintensityvec}
		\varmissintensityvec(\varallocvec) \defeq 
		\left(\varmissintensity_{1}(\varallocation{1}),\ldots,\varmissintensity_{\varnCPs}(\varallocation{\varnCPs}) \right)^{T} .
	\end{equation}  
Finally, we define the overall expected cache miss intensity
	\begin{equation}
		\label{eq:L}
		\varmissintensity(\varallocvec) \defeq
		\sum_{\varCPindex=1}^{\varnCPs}\varmissintensity_{\varCPindex}(\varallocation{\varCPindex}).
	\end{equation}

Motivated by the increasing prevalence of encrypted content delivery, we assume that the operator
cannot observe what content an individual request is for, but it can observe the number of content requests received by a CP and the corresponding number of cache misses. 

Given a static cache partitioning $\varallocvec\in \varfeasibleallocs$, the observed number of content requests and the number of cache misses would form a stationary sequence when measured over subsequent time intervals. The objective of the operator is to find the optimal allocation $\varallocvec^{OPT}$ that minimizes the overall \emph{expected cache miss intensity}, i.e.,
\begin{equation}
\varallocvec^{OPT}\in\arg\min_{\varallocvec\in\varfeasibleallocs}\varmissintensity(\varallocvec),
\end{equation}
based on the measured cache miss intensity. In what follows we propose the \varalgonamelong (\varalgoname) algorithm
that iteratively approximates the optimal allocation.

\begin{table}
\begin{center}
\caption{Frequently used notation (with place of definition)}\label{tab:notation}
	\begin{tabular}{|c|l|}
	\hline 
	$\varnCPs$ & Number of content providers (CP) \tabularnewline
	$\varstorage$ & Available cache slots\tabularnewline
	$\varstorage'$ & Allocated cache slots (\ref{eq:Kprime})\tabularnewline
	$\varallocvec$ & Cache configuration\tabularnewline
    $\varfeasibleallocs$ & Set of feasible cache allocations (\ref{eq:theta})\tabularnewline
    $\varallocationset$ & Set of allowed virtual cache allocations (\ref{eq:C})\tabularnewline
	$\varmissintensity(\varallocvec)$ & Expected cache miss intensity (\ref{eq:L})\tabularnewline
	$\varmissintensityvec$ & Miss intensity vector (\ref{eq:varmissintensityvec})\tabularnewline
	$\varmissintensitylinvec(\varallocvec)$ & Interpolant of the miss intensity (\rlem{uniqueness})\tabularnewline
	$\varallocvec^*$ & Unique minimizer of $\varmissintensitylinvec$ (\rlem{uniqueness})\tabularnewline
	$\varperturbvec\atstep{\vartimeslot}$ & Perturbation vector (\ref{eq:Z})\tabularnewline
	$\vartimeslotlength$ & Time slot length\tabularnewline  
	$\varprojection$ & Euclidean projection (\ref{eq:projection})\tabularnewline  
	$\vargradientvec\atstep{\vartimeslot}$ & Stochastic subgradient  (line~\ref{line:ghat} of \ralg{algo}) \tabularnewline  
	$\varsubgradient(\varallocvec)$ & Subgradient of $\varmissintensitylinvec$ (\ref{eq:barg}) \tabularnewline  
    \hline
	\end{tabular}
\end{center}
\end{table}
\subsection{\varalgonamelong (\varalgoname) Algorithm}
The proposed \varalgoname algorithm is an iterative algorithm that is executed over time slots of fixed duration $\vartimeslotlength$. The pseudo code of the algorithm is shown in \ralg{algo}.
For simplicity we present the algorithm assuming that $\varnCPs$ is even, but the case of an odd number of CPs can be handled by introducing a fictitious CP with zero request rate. 

At time slot $\vartimeslot$ the algorithm maintains a virtual cache allocation $\varallocvec\atstep{\vartimeslot}$.
The virtual allocation is an allocation of  $\varstorage'$ storage slots among the CPs, i.e.,  
	\begin{equation}
		\label{eq:C}
		\varallocvec\atstep{\vartimeslot}\in\varallocationset
		\defeq
		\left\{ \varallocvec\in\mathbb{R}^{p}|\mathbf{1}_{\varnCPs}^{T}\cdot\varallocvec
		\defeq \varstorage' \right\},
	\end{equation}
where
	\begin{equation}
		\label{eq:Kprime}
		\varstorage'=\varstorage-\varnCPs/2.
	\end{equation}
We will justify the introduction of $\varstorage'$ and of $\varallocationset$ in the proof of \rlem{points_generated}. 

\begin{algorithm}[t]
\label{alg:algo}
\small
\caption{\varalgonamelong Algorithm}
Choose an initial allocation $\varallocvec_0\in \varallocationset\cap \varnonnegspace$\\
\For{$k=0$; ; $k++$}{
	Generate $\varperturbvec\atstep{\vartimeslot}$ \\
	$^{+}\varallocvec\atstep{k} = \varcenterptvec(\varallocvec\atstep{k})+\frac{1}{2}\varperturbvec\atstep{\vartimeslot}$ \\
	$^{-}\varallocvec\atstep{k} = \varcenterptvec(\varallocvec\atstep{k})-\frac{1}{2}\varperturbvec\atstep{\vartimeslot}$\\
	Set the configuration to $^{+}\varallocvec\atstep{k}$ for time $\vartimeslotlength/2$ \\
	Measure $^{+}\varmeasuredmissvec\atstep{\vartimeslot}$ \\
	Set the configuration to $^{-}\varallocvec\atstep{k}$ for a time $\vartimeslotlength/2$ \\
	Measure $^{-}\varmeasuredmissvec\atstep{\vartimeslot}$ \\
    $\vardmeasuredmissvec\atstep{k} = ^{+}\varmeasuredmissvec\atstep{\vartimeslot} -  ^{-}\varmeasuredmissvec\atstep{\vartimeslot}$\\
	$\vargradientvec\atstep{k} = \label{line:ghat} \vardmeasuredmissvec\atstep{k} \circ \varperturbvec\atstep{\vartimeslot} - \frac{1}{\varnCPs} \cdot ( {\vardmeasuredmissvec\atstep{k}}^T \cdot \varperturbvec\atstep{\vartimeslot} ) \mathbf{1}_{\varnCPs} $ \\
	$\varallocvec^{(k+1)} = \varprojection( \varallocvec\atstep{k} - \varstepsize\atstep{\vartimeslot}  \vargradientvec\atstep{k} )$\label{line:update}\\
}
\end{algorithm}

In order to obtain from $\varallocvec\atstep{\vartimeslot}$  an integral allocation that can be implemented in the cache,
we define the center-point function $\varcenterptvec:\mathbb{R}^{\varnCPs}\rightarrow\mathbb{R}^{\varnCPs}$,
which assigns to a point in Euclidean space the center of the hypercube containing it, i.e.,
\begin{equation}
	\label{eq:pi}
	\begin{array}{cc}
		\varcenterpt(x)\defeq\lfloor x\rfloor+1/2, & \forall x\in\mathbb{R},\\
		\varcenterptvec(\varallocvec)\defeq\left(\varcenterpt(\varallocation{1}),\ldots,\varcenterpt(\varallocation{\varnCPs})\right)^{T}, 
		& \forall \varallocvec\in\varallocationset,
	\end{array}
\end{equation}
where we use $\lfloor \cdot \rfloor$ to denote the floor of a scalar or of a vector in the component-wise sense.
Furthermore, we define the perturbation vector $\varperturbvec\atstep{\vartimeslot} = (\varperturb\atstep{\vartimeslot}_{1}, \dots,\varperturb\atstep{\vartimeslot}_{\varnCPs})^T$ at time slot $\vartimeslot$,
which is chosen independently and uniform at random from the set of $-1$,$+1$ valued zero-sum vectors
%
\begin{equation}
	\label{eq:Z}
	\varperturbvec\atstep{\vartimeslot} \in \mathcal{Z} \defeq
	\left\{ \mathbf{z}\in\{-1,1\}^{\varnCPs}\left|\mathbf{z}^{T}\cdot\mathbf{1}_{\varnCPs}=0\right.\right\}.
\end{equation}
Given $\varcenterptvec$ and $\varperturbvec\atstep{\vartimeslot}$ the algorithm computes two cache allocations to be implemented during time slot $\vartimeslot$,
	\begin{equation}
		\label{eq:theta_plus_minus}
		\begin{array}{cc}
		^{+}\varallocvec\atstep{\vartimeslot} & \defeq
			\varcenterptvec(\varallocvec\atstep{\vartimeslot}) + \frac{1}{2}\varperturbvec\atstep{\vartimeslot},\\
		^{-}\varallocvec\atstep{\vartimeslot} & \defeq
			\varcenterptvec(\varallocvec\atstep{\vartimeslot}) - \frac{1}{2}\varperturbvec\atstep{\vartimeslot}.
		\end{array}
	\end{equation}
The algorithm first applies allocation $^{+}\varallocvec\atstep{\vartimeslot}$ for  $\vartimeslotlength/2$ amount of time and measures
the cache miss rate $^{+}\varmeasuredmiss\atstep{\vartimeslot}_{\varCPindex}$  for each provider $\varCPindex=1,\ldots,\varnCPs$. It 
then applies allocation $^{-}\varallocvec\atstep{\vartimeslot}$ during the remaining $\vartimeslotlength/2$ amount of time in
slot $\vartimeslot$ and measures the cache miss rates $^{-}\varmeasuredmiss\atstep{\vartimeslot}_{\varCPindex}$. 
The vectors of measured cache misses $^{-}\varmeasuredmissvec\atstep{\vartimeslot} \defeq (^{-}\varmeasuredmiss^{\vartimeslot}_{1},\ldots,^{-}\varmeasuredmiss\atstep{k}_{\varnCPs})^{T}$ and
$^{+}\varmeasuredmissvec\atstep{\vartimeslot} \defeq (^{+}\varmeasuredmiss^{\vartimeslot}_{1},\ldots,^{+}\varmeasuredmiss\atstep{k}_{\varnCPs})^{T}$ are used
to compute the impact $\vardmeasuredmiss\atstep{k}_{\varCPindex} \defeq ^{+}\varmeasuredmiss\atstep{\vartimeslot}_{\varCPindex} - ^{-}\varmeasuredmiss\atstep{\vartimeslot}_{\varCPindex}$ of the perturbation vector
on the cache miss intensity of CP $\varCPindex$, or using the vector notation $\vardmeasuredmissvec\atstep{k} \defeq ^{+}\varmeasuredmissvec\atstep{\vartimeslot} -  ^{-}\varmeasuredmissvec\atstep{\vartimeslot}$.

Based on the measured miss rates, the algorithm then computes the allocation vector $\varallocvec^{(\vartimeslot+1)}$ for the $(k+1)$-th step. Specifically, it first computes (line \ref{line:ghat}, where $\circ$ denotes the Hadamard product) the update vector $\vargradientvec\atstep{\vartimeslot}$, which we show in 
\lcor\ref{cor:stochastic_subgradient} to match in expectation a subgradient of the miss-stream interpolant $\varmissintensitylinvec$, defined in \rlem{uniqueness}. The \mbox{$(k+1)$-th} allocation moves from the $k$-th allocation in the direction of the update vector $\vargradientvec\atstep{\vartimeslot}$,  opportunely scaled by a \emph{step size} $\varstepsize\atstep{k}>0$. Additionally, denoting with $\mathbb{R}_{\geq0}$ the set of non-negative numbers, $\varallocvec^{(\vartimeslot+1)}$ is computed using the Euclidean projection $\varprojection:\varallocationset\rightarrow\varallocationset\cap\mathbb{R}_{\geq0}^{\varnCPs}$, defined as
\begin{equation}
  \label{eq:projection}
	\varprojection(\varallocvec) \defeq
	\arg\min_{\varallocvec^\prime\in\varallocationset\cap\mathbb{R}_{\geq0}^{\varnCPs}}\Vert \varallocvec - \varallocvec^\prime \Vert.
  \end{equation}
\noindent Several remarks are worth making. 
First, we will show in \rlem{projection} that the equation above admits a unique solution and thus the definition is consistent. Second, we will show in \rlem{points_generated} that $\vargradientvec$ computed as in line~\ref{line:ghat} guarantees that the update $\varallocvec\atstep{k} - \varstepsize\atstep{\vartimeslot}  \vargradientvec\atstep{k}$ at line~\ref{line:update} lies inside $\varallocationset$. Nonetheless, this update may have some negative components and we need to project it into $\varallocationset\cap\mathbb{R}_{\geq0}$ by applying $\varprojection$, to ensure that the subsequent virtual allocation $\varallocvec^{(\vartimeslot+1)}$ is valid. 
Third, the \emph{step size} $\varstepsize\atstep{k}$ must be chosen to satisfy $\sum_{k=1}^{\infty} \varstepsize\atstep{k}=\infty$ and $\sum_{k=1}^{\infty} (\varstepsize\atstep{k})^2<\infty$ in order to guarantee convergence (see \rthm{main}). Fourth, although the convergence of the proposed algorithm is guaranteed, for stationary content popularity, irrespectively of the choice of $\varstepsize\atstep{k}$ satisfying the above conditions, we point out that the step size plays an important role in determining the convergence speed, which we will numerically investigate in \lsec\ref{sec:numerical}. 

\section{Convergence Analysis of \varalgoname}\label{sec:analysis}

We first provide definitions and known results (Sec.\ref{sec:preliminaries}) that are instrumental to prove important properties of the proposed algorithm: consistency (Sec.\ref{sec:consistency}), convergence (Sec.\ref{sec:convergence}) and a bound on the optimality gap (Sec.\ref{sec:optimalitygap}). 

\subsection{Preliminaries}\label{sec:preliminaries}
Let us  start by introducing the forward difference defined for functions on discrete sets.
\begin{defn}
\label{def:diff}
	For a function $\mathbf{F}:\mathbb{Z}^{q_1}\rightarrow\mathbb{R}^{q_2}$, $q_1, q_2 \ge 1$
	the forward difference is 
	\[
		\varfwddiff_{n} \mathbf{F}( \mathbf{x} )  \defeq
		 \mathbf{F}( \mathbf{x}+n\cdot \mathbf{1}_{q_1} ) -F(\mathbf{x}), 
		\forall \mathbf{x}\in\mathbb{Z}^{q_1}, n\in \mathbb{Z}\setminus\{0\}.
	\]
	By abuse of notation, we will simply use $\varfwddiff \mathbf{F}( \mathbf{x} )$ to denote $\varfwddiff_1 \mathbf{F}( \mathbf{x} )$.
\end{defn}
The forward difference is convenient for characterizing convexity using the following definition~\cite{Yuceer2002}.
\begin{defn}
\label{def:convex}
	A discrete function $F:\mathbb{Z}\rightarrow\mathbb{R}$ is strictly convex iff $x\rightarrow\varfwddiff F(x)$ is increasing.
\end{defn}
Furthermore, for a class of functions of interest we can establish the following.
\begin{lem}
\label{lem:diff}
Let $F:\mathbb{Z}\rightarrow\mathbb{R}$ decreasing and strictly convex, $x\in\mathbb{Z}$  and $n\in\mathbb{Z}\setminus\{0\}$ we have
	\begin{equation}
	  \varfwddiff_{n}F(x)> n\varfwddiff F(x).\label{eq::zero}
	\end{equation}
\end{lem}
\begin{IEEEproof}

	 We first show that $\forall x,y\in\mathbb{Z}$ such that $y>x$, the following holds
	\begin{eqnarray}
	\varfwddiff_{n}F(y)>\varfwddiff_{n}F(x) & \mathrm{if} \; n>0, \label{eq:first}\\
	\varfwddiff_{n}F(y)<\varfwddiff_{n}F(x) & \mathrm{if} \; n<0. \label{eq:second}
	\end{eqnarray}

	For $n>0$ we can use \rdef{convex} to obtain
	\[
	\varfwddiff_{n}F(y)=\sum_{i=0}^{n-1}\left[F(y+i+1)-F(y+i)\right]=\sum_{i=0}^{n-1}\varfwddiff F(y+i)
	\]
	\begin{equation}
		> \sum_{i=0}^{n-1}\varfwddiff F(x+i)=\varfwddiff_{n}F(x),\label{eq:n_positive}
	\end{equation}
    \noindent which proves (\ref{eq:first}). 

	For $n<0$ algebraic manipulation of the definition of the forward difference and  (\ref{eq:n_positive}) gives
	\[
		\varfwddiff_{n}F(y)=-\varfwddiff_{|n|}F(y-|n|) < -\varfwddiff_{|n|}F(x-|n|)=\varfwddiff_{n}F(x),
	\]
	\noindent which proves (\ref{eq:second}).  To prove (\ref{eq::zero}) for $n>0$, observe that, thanks to \rdef{convex}, each of the $n$ terms of the last summation in (\ref{eq:first}) is lower bounded by $\varfwddiff F(x)$. For $n<0$ via algebraic manipulation we obtain 
	\[
	\varfwddiff_{n}F(x)=-\sum_{i=1}^{|n|}\varfwddiff F(x-i) > -\sum_{i=1}^{|n|}\varfwddiff F(x)=-|n|\cdot\varfwddiff F(x),
	\]
	 which proves (\ref{eq::zero}) as $|n|=-n$.
\end{IEEEproof}
Since \varalgoname generates virtual configurations whose components are not necessarily integer,
we have to extend the discrete functions $\varmissintensity_\varCPindex$ to real numbers.
Thanks to \lthm 2.2 of~\cite{Carnicer1994}, we have the following existence result.
\begin{lem}
\label{lem:convex_extension}
Given a discrete decreasing and strictly convex function $F:\mathbb{Z}\rightarrow\mathbb{R}$, there exists a continuous and strictly convex function $\bar{F}:\mathbb{R}\rightarrow\mathbb{R}$ that extends $F$, i.e., $F(x)=\bar{F}(x),\forall x\in\mathbb{Z}$. We call $\bar{F}$ the \emph{interpolant} of $F$. 
\end{lem}

Finally, we formulate an important property of the Euclidean projection $\varprojection$. 
\begin{lem}
	\label{lem:projection}
	There is a unique function $\varprojection$ satisfying (\ref{eq:projection}). Furthermore, $\varprojection$ satisfies
	\begin{equation}
		\label{eq:no_more_distant_varphi}
		\Vert \varprojection(\varallocvec) - \varallocvec' \Vert \le
		\Vert \varallocvec - \varallocvec' \Vert,
		\forall \varallocvec\in\varallocationset, \varallocvec'\in\varallocationset\cap\varnonnegspace,
	\end{equation}
        i.e., $\varprojection(\varallocvec)$ is no farther from any allocation vector than $\varallocvec$.
\end{lem}
\begin{IEEEproof}
  Observe that $\varallocationset\cap\varnonnegspace$ is a simplex, and thus closed and convex.  Hence, the Euclidean projection
  $\varprojection$ is the unique solution of (\ref{eq:projection})~\cite{Wang2013g}. 
  Furthermore, the Euclidean projection is non-expansive (see, e.g., Fact 1.5 in \cite{Bauschke1996}), i.e., for $\varallocvec,\varallocvec'\in\varallocationset$ it satisfies
  $
    \Vert \varprojection(\varallocvec) - \varprojection(\varallocvec') \Vert \leq \Vert \varallocvec - \varallocvec' \Vert.
   $
   Observing that if $\varallocvec'\in\varallocationset\cap\varnonnegspace$ then $\varprojection(\varallocvec')=\varallocvec'$ proves the result.
\end{IEEEproof}

\subsection{Consistency}\label{sec:consistency}
We first have to prove that during each time slot the configurations $^{-}\varallocvec\atstep{k},^{+}\varallocvec\atstep{k}$ that \varalgoname imposes on the cache are feasible.
This is non-trivial, as the operators used in computing the allocations are defined on proper subsets of $\mathbb{R}^p$. The following lemma establishes that the
allocations computed by \varalgoname always fall into these subsets.
\begin{lem}
	\label{lem:points_generated}
	The allocations $\varallocvec\atstep{\vartimeslot}$ are consistent in every time slot, as they satisfy
	\begin{enumerate}[(a)]
		\item
			$\varallocvec\atstep{k} - a_k  \vargradientvec\atstep{k} \in \varallocationset$,
		\item
			$\varallocvec^{(\vartimeslot+1)}\in\varallocationset\cap\varnonnegspace$,
		\item
			$^{+}\varallocvec\atstep{\vartimeslot},^{-}\varallocvec\atstep{\vartimeslot}\in\varfeasibleallocs$.
	\end{enumerate}
\end{lem}
\begin{IEEEproof}
Recall that $\varallocvec_0\in\varallocationset\cap\varnonnegspace$. To show (a) observe that 
\begin{equation}
	\label{eq:ghat}
	\vargradientvec\atstep{\vartimeslot} \cdot \mathbf{1}_{\varnCPs}=\sum_{j=1}^{\varnCPs}\delta \varmeasuredmiss\atstep{\vartimeslot}_{\varCPindex}\cdot\varperturb\atstep{\vartimeslot}_{\varCPindex}-\sum_{j=1}^{\varnCPs}\delta \varmeasuredmiss\atstep{\vartimeslot}_{\varCPindex}\cdot\varperturb\atstep{\vartimeslot}_{\varCPindex}=0,
\end{equation}
and thus if $\varallocvec\atstep{\vartimeslot}\in\varallocationset$, then $\varallocvec\atstep{\vartimeslot} - \varstepsize\atstep{\vartimeslot}  \vargradientvec\atstep{\vartimeslot}\in\varallocationset$.
The definition of the Euclidean projection (\ref{eq:projection}) and (a) together imply (b). Finally, observe that 
	\begin{equation}
		\mathbf{1} \cdot ^{+}\varallocvec\atstep{\vartimeslot} = \mathbf{1} \cdot \lfloor \varallocvec \rfloor + \frac{\varnCPs}{2}
		\le \mathbf{1} \cdot  \varallocvec  + \frac{\varnCPs}{2} \le \varstorage^\prime+\frac{\varnCPs}{2}=\varstorage,
	\end{equation}
        which proves (c). Note that the above motivates the choice of $\varstorage^\prime$ in the definition of the set of virtual allocations $\varallocationset$,
        as if $\varstorage^\prime>\varstorage-\frac{\varnCPs}{2}$ then $^{+}\varallocvec\atstep{\vartimeslot},^{-}\varallocvec\atstep{\vartimeslot}\in\varfeasibleallocs$ may be violated due to
        the use of the mapping $\varcenterpt$ and $\varperturbvec\atstep{\vartimeslot}$ in (\ref{eq:theta_plus_minus}).       
\end{IEEEproof}

\subsection{Convergence}\label{sec:convergence}

To prove convergence of \varalgoname, we first consider the relationship between the measured miss rates $^{+}\varmeasuredmiss\atstep{\vartimeslot}_{\varCPindex}$
and $^{-}\varmeasuredmiss\atstep{\vartimeslot}_{\varCPindex}$ and the expected miss intensities $\varmissintensity_{\varCPindex}(^{+}\varallocation{\varCPindex}\atstep{\vartimeslot})$ and
$\varmissintensity_{\varCPindex}(^{-}\varallocation{\varCPindex}\atstep{\vartimeslot})$, respectively.
We define the measurement noise
\begin{equation}
	\label{eq:noise}
	\begin{array}{cc}
	^{+}\varnoisevec\atstep{\vartimeslot} & \defeq  ^{+}\varmeasuredmissvec\atstep{\vartimeslot} -
		\varmissintensityvec(^{+}\varallocvec\atstep{\vartimeslot}),\\
	^{-}\varnoisevec\atstep{\vartimeslot} & \defeq ^{-}\varmeasuredmissvec\atstep{\vartimeslot} -
		\varmissintensityvec(^{-}\varallocvec\atstep{\vartimeslot})
	\end{array}
\end{equation}
and the corresponding differences 
\begin{equation}
	\label{eq:delta_small}
	\begin{array}{cc}
		\delta \varnoisevec\atstep{\vartimeslot} \defeq & ^{+}\varnoisevec\atstep{\vartimeslot} - ^{-}\varnoisevec\atstep{\vartimeslot},\\
		\delta \varmissintensityvec\atstep{\vartimeslot} \defeq & \varmissintensityvec(^{+}\varallocvec\atstep{\vartimeslot}) - 
			\varmissintensityvec(^{-}\varallocvec\atstep{\vartimeslot}).\\
	\end{array}
\end{equation}
Observe that $\varperturbvec\atstep{\vartimeslot}$, $^{+}\varmeasuredmissvec\atstep{\vartimeslot}$ and $^{-}\varmeasuredmissvec\atstep{\vartimeslot}$
are random variables and form a stochastic process.
Using these definitions we can formulate the following statement about the measured miss rates.
\begin{lem}
\label{lem:noise_zero}
	The conditional expectation of the measurement noise and its difference satisfy
	\begin{equation}
		\expectation[\delta\varnoisevec\atstep{\vartimeslot} | \varallocvec\atstep{\vartimeslot}] = \mathbf{0}_p.
	\end{equation}
\end{lem}
\begin{IEEEproof}
  Observe that due to the stationarity of the request arrival processes we have $\expectation[^{+}\varnoisevec\atstep{\vartimeslot}| \varallocvec\atstep{\vartimeslot}]=0$ and $\expectation[^{-}\varnoisevec\atstep{\vartimeslot}| \varallocvec\atstep{\vartimeslot}]=0$,
  which due to the additive law of expectation yields the result. 
  \end{IEEEproof}
Intuitively, this is equivalent to saying that 
the sample averages provide an unbiased estimator of the miss rates. In what follows we establish an analogous result for the update vector $\vargradientvec\atstep{\vartimeslot}$ with respect to a subgradient of the interpolant $\varmissintensitylinvec$ of the expected miss intensity $\varmissintensity$, which itself is a discrete function. We define and characterize $\varmissintensitylinvec$ in the following lemma, which recalls known results from convex optimization.
\begin{lem}
\label{lem:uniqueness}
Given the interpolants $\varmissintensitylinvec_{\varCPindex}$ of the expected miss intensities $\varmissintensity_{\varCPindex}$ of the CPs and defining the 
interpolant of $\varmissintensity$ as 
$\varmissintensitylinvec(\varallocvec)\defeq
	\sum_{\varCPindex=1}^{\varnCPs} 
		\varmissintensitylinvec_{\varCPindex}(\varallocation{\varCPindex}),
	\forall\mathbf{\varallocvec}\in\mathbb{R}_{\geq 0}^{\varnCPs}
$, $\varmissintensitylinvec$ is strictly convex and admits a unique minimizer $\varallocvec^*$ in $\varallocationset\cap\varnonnegspace$.
\end{lem}
\begin{IEEEproof}
Recall that each interpolant $\varmissintensitylinvec_{\varCPindex}$ of $\varmissintensity_{\varCPindex}$ is strictly convex as shown in \rlem{convex_extension}.
The strict convexity of $\varmissintensitylinvec$ can then be obtained applying \lthm 1.17 of~\cite{Sefanov2013}. Then, we observe that $\varallocvec^*$ is the solution
to a convex optimization problem with a strictly convex objective function, which is unique (\lsec 4.2.1 of~\cite{Boyd2010}. 
\end{IEEEproof}
For completeness, let us recall the definition of a subgradient of a function from (see, e.g.,~\cite{Shor1998}).
\begin{defn}
  \label{def:subgradient}
  Given a function $\bar{\varmissintensity}:\mathbb{R}^{p}\rightarrow\mathbb{R}$, a function $\varsubgradient:\varallocationset\subseteq\mathbb{R}^{\varnCPs}\rightarrow\mathbb{R}^{\varnCPs}$
is a subgradient of $\bar{\varmissintensity}$ over $\varallocationset$ iff 
\[
	\bar{\varmissintensity}(\varallocvec^\prime)-\bar{\varmissintensity}(\varallocvec)
	\ge\varsubgradient(\varallocvec)^{T}
	\cdot(\varallocvec^\prime-\varallocvec),
	\forall \varallocvec,\varallocvec^\prime\in\varallocationset.
\]
\end{defn}
We are now ready to introduce a subgradient $\varsubgradient(\varallocvec)$ for the interpolant of the expected cache miss intensity $\varmissintensitylinvec$.
\begin{lem}
	\label{lem:extension}
        The function
\begin{equation}
	\varsubgradient(\varallocvec) \defeq
	\varfwddiff\varmissintensityvec\atstep{\vartimeslot}(\lfloor\varallocvec\rfloor)-
	\frac{1}{\varnCPs} \cdot \varfwddiff L(\lfloor\varallocvec\rfloor) \cdot \mathbf{1}_{\varnCPs}\label{eq:barg}
\end{equation}
is a subgradient of $\varmissintensitylinvec$ over $\varallocationset\cap\varnonnegspace$.
\end{lem}
\begin{IEEEproof}
Observe that for  $\varallocvec, \varallocvec^\prime\in\varallocationset$ 
\[
	\varsubgradient(\varallocvec)^{T}\cdot(\varallocvec'-\varallocvec)=
	\varfwddiff\varmissintensityvec\atstep{\vartimeslot}(\lfloor\varallocvec\rfloor) ^T
	\cdot(\varallocvec'-\varallocvec)
\]
\[
	-\frac{1}{\varnCPs} \cdot 
	\varfwddiff L(\lfloor\varallocvec\rfloor)	
	\cdot\left[\mathbf{1}_{\varnCPs}^{T}\cdot(\varallocvec'-\varallocvec)\right]. 
\]
At the same time, for $\varallocvec, \varallocvec^\prime\in\varallocationset$ we have 
\[
	\mathbf{1}_{\varnCPs}^{T}\cdot(\varallocvec'-\varallocvec)=
	(\mathbf{1}_{\varnCPs}^{T}\cdot\varallocvec'-\mathbf{1}_{\varnCPs}^{T}\cdot\varallocvec)=\varstorage^\prime-\varstorage^\prime=0.
\]
Therefore, for any $\varallocvec,\varallocvec'\in\varallocationset$
\begin{equation}
	\label{eq:uela}
	\varsubgradient(\varallocvec)^{T}\cdot(\varallocvec'-\mathbf{\varallocvec} )=
	\varfwddiff\varmissintensityvec\atstep{\vartimeslot}(\lfloor\varallocvec\rfloor)
	\cdot(\varallocvec'-\varallocvec ).
\end{equation}
Thus, according to \rdef{subgradient}, in order to show that $\varsubgradient$ is a subgradient of $\varmissintensitylinvec$ it suffices to show that
\begin{equation}
	\label{eq:goal}
	\sum_{\varCPindex=1}^{\varnCPs}\left[\varmissintensitylinvec_{\varCPindex}(\theta_{\varCPindex}')-\varmissintensitylinvec_{\varCPindex}(\theta_{\varCPindex})\right]
	\ge\sum_{\varCPindex=1}^{\varnCPs} \varfwddiff \varmissintensity_{\varCPindex} (\lfloor \theta_j \rfloor) \cdot (\theta_{\varCPindex}'-\theta_{\varCPindex}).
\end{equation}
We now show that this holds component-wise. If \mbox{$\lfloor\theta_{\varCPindex}'\rfloor-\lfloor\theta_{\varCPindex}\rfloor=0$}, then the above clearly holds.
Otherwise, if \mbox{$n=\lfloor\theta_{\varCPindex}'\rfloor-\lfloor\theta_{\varCPindex}\rfloor\neq 0$} we apply a well known property of convex functions (\lthm 1.3.1 of \cite{Niculescu2004}) to obtain:
\begin{eqnarray}
	\frac{\varmissintensitylinvec_{\varCPindex}(\lfloor\varallocation{\varCPindex}'\rfloor)-\varmissintensitylinvec_{\varCPindex}(\lfloor\varallocation{\varCPindex}\rfloor)}{(\lfloor\varallocation{\varCPindex}'\rfloor-\lfloor\varallocation{\varCPindex}\rfloor)}
	\le
	\frac{\varmissintensitylinvec_{\varCPindex}(\varallocation{\varCPindex}')-\varmissintensitylinvec_{\varCPindex}(\varallocation{\varCPindex})}{(\varallocation{\varCPindex}'-\varallocation{\varCPindex})}
	\nonumber
	\\ \le
	\frac{\varmissintensitylinvec_{\varCPindex}(\lfloor\varallocation{\varCPindex}'\rfloor+1)-\varmissintensitylinvec_{\varCPindex}(\lfloor\varallocation{\varCPindex}\rfloor+1)}{(\lfloor\varallocation{\varCPindex}'\rfloor+1-(\lfloor\varallocation{\varCPindex}\rfloor+1))},
	\nonumber
\end{eqnarray}
which, by \rdef{diff}, can be rewritten as:
\begin{equation}
	\label{eq:chain}
	\frac{ \varfwddiff_n L_{j}(\lfloor\theta_{j}\rfloor) }{n}
	\le
	\frac{\varmissintensitylinvec_{\varCPindex}(\varallocation{\varCPindex}')-\varmissintensitylinvec_{\varCPindex}(\varallocation{\varCPindex})}{\varallocation{\varCPindex}'-\varallocation{\varCPindex}}
	\le
	\frac{ \varfwddiff_n L_{j}(\lfloor\theta_{j}+1 \rfloor) }{n}.
\end{equation}
For $n>0$ we can use the first inequality of (\ref{eq:chain}) and \rlem{diff} to obtain
\begin{equation}
	\label{eq:componentwise_ineq}
	\varmissintensitylinvec_{\varCPindex}(\varallocation{\varCPindex}')-
	\varmissintensitylinvec_{\varCPindex}(\varallocation{\varCPindex})
	\ge
	\varfwddiff L_{j}(\lfloor\theta_{j}\rfloor) \cdot (\varallocation{\varCPindex}'-\varallocation{\varCPindex}).
\end{equation}
For $n<0$ we can use  the second inequality of (\ref{eq:chain}) and \rlem{diff} to obtain
\begin{equation}
	\label{eq:chain2}
	\frac{\varmissintensitylinvec_{\varCPindex}(\varallocation{\varCPindex}')-\varmissintensitylinvec_{\varCPindex}(\varallocation{\varCPindex})}
	{\varallocation{\varCPindex}'-\varallocation{\varCPindex}}
	\le
	\varfwddiff L_{j}(\lfloor\theta_{j}+1 \rfloor)
	\le
	\varfwddiff L_{j}(\lfloor\theta_{j} \rfloor)
\end{equation}
and by multiplying the first and the second term of (\ref{eq:chain2}) by $\varallocation{\varCPindex}'-\varallocation{\varCPindex}$  (which is negative since $n=\lfloor\theta_{\varCPindex}'\rfloor-\lfloor\theta_{\varCPindex}\rfloor$ is negative), we obtain the result. 

\end{IEEEproof}
The subgradient $\varsubgradient$ will be central to proving the convergence of \varalgoname, but it cannot be measured directly.
The next proposition establishes a link between the update vector $\vargradientvec\atstep{\vartimeslot}$, which we compute in every time slot, and
the subgradient $\varsubgradient$.
\begin{prop}
\label{prop:ghat}
The update vector $\vargradientvec\atstep{\vartimeslot}$ is composed of the subgradient $\varsubgradient$ plus a component due to the noise, 
\[
	\vargradientvec\atstep{\vartimeslot}=
	\varsubgradient(\varallocvec\atstep{\vartimeslot})+\delta\varnoisevec\atstep{\vartimeslot}\circ\varperturbvec\atstep{\vartimeslot}-
	\frac{1}{\varnCPs} \cdot \left[{\delta\varnoisevec\atstep{\vartimeslot}}^{T}\cdot\varperturbvec\atstep{\vartimeslot}\right]\mathbf{1}_{\varnCPs} . 
\]
\end{prop}
\begin{IEEEproof}
We first apply (\ref{eq:delta_small}) to obtain
\begin{eqnarray}
  \label{eq::ghat-expanded}
		\vargradientvec\atstep{\vartimeslot}&=&
		\delta\varmissintensityvec\atstep{\vartimeslot}\circ\varperturbvec\atstep{\vartimeslot}-
		\frac{1}{\varnCPs} \cdot ({\delta\varmissintensityvec\atstep{\vartimeslot}}^{T}\cdot\varperturbvec\atstep{\vartimeslot})\mathbf{1}_{\varnCPs}\\
		&+&\delta\varnoisevec\atstep{\vartimeslot}\circ\varperturbvec\atstep{\vartimeslot}-
		\frac{1}{\varnCPs} \cdot ({\delta\varnoisevec\atstep{\vartimeslot}}^{T}\cdot\varperturbvec\atstep{\vartimeslot})\mathbf{1}_{\varnCPs}.\nonumber
	\end{eqnarray}
	Consider now a particular realization of the random variable $\varperturbvec\atstep{\vartimeslot}$. We can express component $\varCPindex$ of $\delta\varmissintensityvec\atstep{\vartimeslot}\circ\varperturbvec\atstep{\vartimeslot}=\left[\varmissintensityvec(^{+}\varallocvec\atstep{k})-\varmissintensityvec(^{-}\varallocvec\atstep{\vartimeslot})\right]\circ\varperturbvec\atstep{\vartimeslot}$
	as
	\[
	\left[\varmissintensity_{\varCPindex}\left(\Pi\left(\varallocation{\varCPindex}\atstep{\vartimeslot}\right)+\frac{1}{2}\varperturb\atstep{\vartimeslot}_{\varCPindex}\right)-\varmissintensity_{\varCPindex}\left(\Pi\left(\theta\atstep{\vartimeslot}_{\varCPindex}\right)-\frac{1}{2}\varperturb\atstep{\vartimeslot}_{\varCPindex}\right)\right]
	\]
	\[
	\cdot\varperturb\atstep{\vartimeslot}_{\varCPindex}
	\]
	\[
	=\left[\varmissintensity_{\varCPindex}\left(\Pi\left(\varallocation{\varCPindex}\atstep{\vartimeslot}\right)+\frac{1}{2}\right)-\varmissintensity_{\varCPindex}\left(\Pi\left(\varallocation{\varCPindex}\atstep{\vartimeslot}\right)-\frac{1}{2}\right)\right]
	\]
	\[
	=\left[\varmissintensity_{\varCPindex}\left(\lfloor\varallocation{\varCPindex}\atstep{\vartimeslot}\rfloor+1\right)-\varmissintensity_{\varCPindex}\left(\lfloor\varallocation{\varCPindex}\atstep{\vartimeslot}\rfloor\right)\right],
	\]
	 where the first equality can be easily verified assuming that $\varperturb\atstep{\vartimeslot}_{\varCPindex}=-1$ and then assuming that it is $\varperturb\atstep{\vartimeslot}_{\varCPindex}=1$. We thus obtain 
	\[
		\delta\varmissintensityvec\atstep{\vartimeslot}\circ\varperturbvec\atstep{\vartimeslot}=
		\varfwddiff \varmissintensityvec(\lfloor\varallocvec\atstep{\vartimeslot}\rfloor)
                \]
                and in scalar form
                \[
		{\delta\varmissintensityvec\atstep{\vartimeslot}}^{T}\cdot\varperturbvec\atstep{\vartimeslot}=
		\varfwddiff \varmissintensity(\lfloor\varallocvec\atstep{\vartimeslot}\rfloor).
	\]

By substituting these in (\ref{eq::ghat-expanded})
	\[
		\vargradientvec\atstep{\vartimeslot} 
		= \varfwddiff\varmissintensityvec(\lfloor\varallocvec\atstep{\vartimeslot}\rfloor)
		-\frac{1}{\varnCPs} \cdot 
		\varfwddiff \varmissintensity(\lfloor\varallocvec\atstep{\vartimeslot}\rfloor) \cdot 
		\mathbf{1}_{\varnCPs}
	\]
	\[
		+\delta\varnoisevec\atstep{\vartimeslot}\circ\varperturbvec\atstep{\vartimeslot}-
		\frac{1}{\varnCPs} \cdot ({\delta\varnoisevec\atstep{\vartimeslot}}^{T}\cdot\varperturbvec\atstep{\vartimeslot})\mathbf{1}_{\varnCPs}
	\]
and using (\ref{eq:barg}), we obtain the result.
\end{IEEEproof}

\noindent Furthermore, thanks to \rlem{noise_zero}, the second term of (\ref{eq::ghat-expanded}), which is due to the noise, is zero in expectation, which provides
the link between the update vector $\vargradientvec\atstep{\vartimeslot}$ and the subgradient $\varsubgradient(\varallocvec\atstep{k})$.
\begin{corollary}
	\label{cor:stochastic_subgradient} 
	The conditional expectation of $\vargradientvec\atstep{\vartimeslot}$ is 
$\expectation[\vargradientvec\atstep{\vartimeslot}| \varallocvec\atstep{\vartimeslot}]=\varsubgradient(\varallocvec\atstep{k})$
 and thus $\vargradientvec\atstep{\vartimeslot}$ is a stochastic subgradient of $\varmissintensitylinvec$, i.e. $\expectation[\vargradientvec\atstep{\vartimeslot}]=\varsubgradient(\varallocvec\atstep{k})$.
\end{corollary}

\noindent This leads us to the following theorem. 
\begin{thm}
	\label{thm:main}
	The sequence $\varallocvec\atstep{k}$ generated by \varalgoname converges in probability to the unique minimizer $\varallocvec^*$of $\varmissintensitylinvec$, i.e., for arbitrary $\delta>0$
	 $$\lim_{k\rightarrow\infty} Pr\{ \Vert \varallocvec\atstep{k} - \varallocvec^*\Vert > \delta\}= 0.$$
\end{thm}
\begin{IEEEproof}
 
   The proof of convergence is similar to (\lthm 46 in~\cite{Shor1998}), with the difference
   that our proof holds for Euclidean projection-based stochastic subgradients.
   Let us compute
\begin{eqnarray}
	\nonumber
	\lefteqn{\Vert \varallocvec\atstep{\vartimeslot+1} - \varallocvec^* \Vert^2=
	\Vert \varprojection( \varallocvec\atstep{\vartimeslot} - \varstepsize\atstep{\vartimeslot}  \vargradientvec\atstep{\vartimeslot} ) - 
		\varallocvec^* \Vert^2
	}\\ 
	\nonumber
	&\le&
	\Vert \varallocvec\atstep{\vartimeslot} - \varstepsize\atstep{\vartimeslot}  \vargradientvec\atstep{\vartimeslot} - 
		\varallocvec^* \Vert^2\\
	\nonumber
	\label{eq:normcalculus}
	&=& \Vert \varallocvec\atstep{\vartimeslot} - \varallocvec^* \Vert^2 - 
		2  \varstepsize\atstep{\vartimeslot} \cdot (\vargradientvec\atstep{\vartimeslot} )^T \cdot (\varallocvec\atstep{\vartimeslot} - \varallocvec^*)\\
		&&+ (\varstepsize\atstep{\vartimeslot})^2 \cdot \Vert \vargradientvec\atstep{\vartimeslot} \Vert^2,
\end{eqnarray}
where the first inequality is due to \rlem{projection}. Thanks to 
cor~\ref{cor:stochastic_subgradient} and \rdef{subgradient}
\[
	\left(\expectation[\vargradientvec\atstep{\vartimeslot}\vert\varallocvec\atstep{\vartimeslot}]\right)^T
	\cdot (\varallocvec\atstep{\vartimeslot} - \varallocvec^*) = 
	\left( \varsubgradient(\varallocvec\atstep{\vartimeslot}) \right) ^T \cdot (\varallocvec\atstep{\vartimeslot} - \varallocvec^*) \ge 0.
\]
  Recall that the number of arriving requests per time slot $\vararrivals(\vartimeslotlength)$ is bounded, and thus $\Vert \vargradientvec\atstep{\vartimeslot}\Vert^2$ is bounded, i.e., $\Vert \vargradientvec\atstep{\vartimeslot}\Vert^2\leq\varuniformbound$ for some $0<c<\infty$.  Hence, applying the expectation to (\ref{eq:normcalculus})
\begin{eqnarray}
	\expectation\left[ \left. \Vert \varallocvec^{(\vartimeslot+1)} - \varallocvec^* \Vert^2 \right| \varallocvec\atstep{\vartimeslot} \right]
	\le
	\Vert \varallocvec\atstep{\vartimeslot} - \varallocvec^* \Vert^2 + \varuniformbound (\varstepsize\atstep{\vartimeslot})^2. \label{eq:lastinequality}
\end{eqnarray}

 
 Defining the random variable $$\mbox{$z_\vartimeslot \defeq \Vert \varallocvec\atstep{\vartimeslot} - \varallocvec^* \Vert^2 + \varuniformbound \sum_{s=\vartimeslot}^\infty(\varstepsize^{(s)})^2$},$$ it can be easily verified that (\ref{eq:lastinequality}) is equivalent to the inequality 
$\expectation[ z_{\vartimeslot+1} \vert z_\vartimeslot,\dots,z_1 ]\le z_\vartimeslot$.
Consequently, $\{z_\vartimeslot\}_{\vartimeslot=1}^\infty$ is a supermartingale and converges almost surely to a limit  $z^*$. 
Recalling now one of the required properties of the step size sequence, i.e.,  $\lim_{\vartimeslot\rightarrow \infty} \sum_{s=\vartimeslot}^\infty (\varstepsize\atstep{\vartimeslot})^2=0$, we have that the sequence $\{ \Vert \varallocvec\atstep{\vartimeslot} - \varallocvec^* \Vert^2 \}$ also converges to $z^*$ with probability one.


We now show by contradiction that the limit $z^*$ is equal to zero. If this were not true, then one could
find $\epsilon>0$ and $\delta>0$ such that, with probability $\delta>0$, $\Vert \varallocvec\atstep{\vartimeslot} - \varallocvec^* \Vert\ge \epsilon$ for all sufficiently large $\vartimeslot$, and thus
\[
	\sum_{\vartimeslot=0}^\infty \varstepsize\atstep{\vartimeslot} \cdot 
		(\expectation[\vargradientvec\atstep{\vartimeslot} | \varallocvec\atstep{\vartimeslot}])^T \cdot (\varallocvec\atstep{\vartimeslot} - \varallocvec^*)
	=+\infty,
\]
with probability $\delta$, which would imply
\[
	\expectation\left[ \sum_{\vartimeslot=0}^\infty \varstepsize\atstep{\vartimeslot} 
	\cdot 
	\left(\varsubgradient(\varallocvec\atstep{k}) \right)^T \cdot (\varallocvec\atstep{\vartimeslot} - \varallocvec^*)
	\right]
	=+\infty.
\]
However, this would contradict the following relation (which is obtained by a recursion on (\ref{eq:normcalculus}) and then applying the expectation)
\begin{eqnarray}
	\nonumber
	\lefteqn{\expectation [ \Vert \varallocvec^{(\vartimeslot+1)} - \varallocvec^* \Vert^2]\leq} \\
	\nonumber
	&&\Vert \varallocvec^{(0)} - \varallocvec^* \Vert^2
	-2 \expectation \left[ \sum_{s=0}^\vartimeslot \varstepsize^{(s)} \cdot (\vargradientvec^{(s)})^T \cdot (\varallocvec^{(s)} - \varallocvec^*) \right]
	+ \\
	\nonumber&&
	\expectation \left[ \sum_{s=0}^\vartimeslot  \varstepsize^{(s)}\cdot \Vert \vargradientvec^{(s)}) \Vert^2 \right],
\end{eqnarray}
as the left hand side cannot be negative.
 
\end{IEEEproof}

\subsection{Optimality gap}\label{sec:optimalitygap}

It is worthwhile to note that the minimizer $\varallocvec^*$ of $\varmissintensitylinvec$ over $\varallocationset\cap\varnonnegspace$ may not coincide with its minimizer $\varallocvec^{OPT}$ over $\varfeasibleallocs$ for two reasons: i) $\varstorage^{\prime}<\varstorage$ and ii) $\varallocvec^{OPT}$ is forced to have integer components while $\varallocvec^*$ is can be a real vector. In what follows we show that the optimality gap $\Vert \varallocvec^{OPT} - \varallocvec^* \Vert_\infty$ is bounded by a small number, compared to the number of cache slots available.

\begin{lem}
The gap between the optimal solution $\varallocvec^{OPT}$ and the configuration $\varallocvec^*$ to which \varalgoname converges is $\Vert\varallocvec^*-\varallocvec^{OPT}\Vert_{\infty}\leq(3/2)\varnCPs$
\end{lem}
\begin{IEEEproof}
We observe that $\varallocvec^*$ is the optimal solution of the continuous Simple Allocation Problem (SAP), expressed as 
$\max \left( - \sum_{\varCPindex=1}^\varnCPs \varmissintensity_\varCPindex (\varallocation{\varCPindex}) \right)$, subject to $\sum_{\varCPindex=1}^\varnCPs \varallocation{\varCPindex} \le \varstorage'$ with $\varallocvec\in\varnonnegspace$. $\varstorage'$ usually referred to as \emph{volume} and we denote with $\varsapcont{\varstorage'}$ the problem above. The integer version of the SAP, which we denote by $\varsapint{\varstorage'}$, is obtained from the problem above with the additional constraint $\varallocvec\in\mathbb{Z}^p$.
According to \rcor~4.3 of~\cite{Hochbaum1994} there exists a solution $\varallocdiscr$ of $\varsapint{\varstorage'}$ such that $\Vert\varallocvec^*-\varallocdiscr \Vert_{\infty}\leq\varnCPs$. The solution of the integer SAP can be constructed via the greedy algorithm presented in \lsec 2 of \cite{Hochbaum1994}. In our case, it consists of iteratively adding storage slots, one by one, each time to the CP whose miss intensity is decreased the most by using this additional slot. Based on this, it is easy to verify that a solution $\varallocvec^{OPT}$ can be obtained starting from $\varallocdiscr$ and adding the remaining $\varstorage-\varstorage'$ slots. Therefore, $\Vert \varallocvec^{OPT}-\varallocdiscr \Vert_\infty \le \varnCPs/2$, which implies
\[
	\Vert \varallocvec^{OPT} - \varallocvec^* \Vert_\infty \le
	\Vert \varallocvec^* - \varallocdiscr \Vert_\infty + 
	\Vert \varallocvec^{OPT} - \varallocdiscr \Vert_\infty \le
	(3/2)\varnCPs.
\]
\end{IEEEproof}


\section{Performance evaluation}\label{sec:numerical}
We evaluate the performance of \varalgoname through simulations performed in Octave. We first describe the evaluation scenario (Sec.\ref{res:scen}) and show how the convergence speed is impacted by the choice of the step size sequence (Sec.\ref{res:conv}). We then evaluate the sensitivity of \varalgoname to various system parameters (Sec.\ref{res:sens}).
Finally, recognizing that content catalogs are rarely static in the real world, we investigate the expected performance in the case of changing content catalogs (Sec.\ref{res:nonstat}).

\subsection{Evaluation Scenario}\label{res:scen}

\renewcommand\mycommfont[1]{\emph{#1}}
\SetCommentSty{mycommfont}

\begin{algorithm}[t]
\label{alg:conditional}
\small
\caption{Conditional Step Size Sequence Computation}
$\varstepsize = \frac{p}{\Vert \vargradientvec^{(1)} \Vert_1} \cdot \frac{\varstorage'}{p}$\\
$b=a/10$\\
\uIf{$\vartimeslot\le \vartimeslot_{BS}$ \tcp*{Bootstrap Phase}}{
	
	$\varstepsize\atstep{\vartimeslot} = \varstepsize$
}
\uElseIf{$\vartimeslot \le M$ \tcp*{Adaptive Phase}}{
	Compute the miss-ratio $\varmissratio\atstep{\vartimeslot}$ 
		during the current iteration \\
	Compute $\varmissratio_{5th}$, i.e. the $5$th percentile of the previous miss ratios $\varmissratio^{(1)},\dots,\varmissratio^{(k-1)}$\\
	$\hat{\varstepsize}\atstep{\vartimeslot}=\varstepsize^{(\vartimeslot-1)}/2$\\
	$\tilde{\varstepsize}\atstep{\vartimeslot}=
			\varstepsize^{(\vartimeslot-1)} - 
			\frac{\varstepsize^{(\vartimeslot-1)} - b}
				{M-\vartimeslot+1}
	$\\
	$\varstepsize\atstep{\vartimeslot} =
		\begin{cases}
		 \left( \min(\hat{\varstepsize}\atstep{\vartimeslot}, 
			\tilde{\varstepsize}\atstep{\vartimeslot}),
			b
		 \right) & 
 		 \varmissratio\atstep{\vartimeslot}\le \varmissratio_{5th} \\
		 \tilde{\varstepsize}\atstep{\vartimeslot} &
		 \mbox{otherwise}
		\end{cases}
	$
}
	\Else{	 
	$\varstepsize\atstep{\vartimeslot} =
		\varstepsize^{(\vartimeslot-1)} \cdot
		\left(
			1-\frac{1}{(1+\vartimeslot)}
		\right)^{\frac{1}{2}+\varconditionalexp}
	$	\tcp*{Moderate Phase}
	}

\end{algorithm}

We consider a content catalog of $10^8$ objects, in line with the literature~\cite{Fricker2012}\censure{\cite{Ardelius2012}} and recent measurements~\cite{Imbrenda2014}.
We partition the catalog in disjoint sub-catalogs, one per each CP. We assume that the content popularity in each sub-catalog follows Zipf's law with exponent $\alpha=0.8$, as usually done in the literature~\cite{Poularakis2015}\censure{\cite{Fricker2012}moreover the ref is broken}. We use a cache size of $K \in \{ 10^4,10^5,10^6\}$ objects (which corresponds to cache/catalog ratios of $10^{-4},10^{-3}$ and $10^{-2}$ respectively). In practice, the request arrival rate may depend on several factors such as the cache placement in the network hierarchy, the level of aggregation, the time of day, etc.  Without loss of generality, we set the request arrival rate to $\lambda=10^{2}\mbox{req/s}$, according to recent measurements performed on ISP access networks~\cite{Imbrenda2014}.
We compare the performance of  \varalgoname to that of the optimal allocation $\varallocvec^{OPT}$ (\emph{Opt}), and to that of the naive solution in which the cache space $K$ is equally divided among all the CPs and is unchanged throughout the simulation (\emph{Unif}).

While we proved convergence of \varalgoname, the speed of convergence is crucial to let the algorithm also be of practical use: we thus consider three step size sequences, as follows.
In the \emph{Reciprocal} scheme, the step size is $\varstepsize\atstep{\vartimeslot}=a/\vartimeslot$, where $\varstepsize=\frac{1}{\Vert \vargradientvec^{(1)} \Vert}\cdot\frac{\varstorage'}{p}$.
Observe that, with this choice, the Euclidean norm of the first update $\varstepsize\atstep{1}\cdot\vargradientvec\atstep{1} $ is $\frac{\varstorage'}{p}$, which allows to change this amount of slots in the allocation, thus obtaining a broad exploration of the allocation space at the very beginning.

In the \emph{Moderate} scheme, step sizes decrease slowly, to avoid confining the exploration only to the beginning. We resort to guidelines of \cite{Wang2013f}, and define the step size as
$\varstepsize\atstep{\vartimeslot} =
		\varstepsize^{(\vartimeslot-1)} \cdot
		\left(
			1-\frac{1}{(1+M+\vartimeslot)}
		\right)^{\frac{1}{2}+\varconditionalexp}
	$	
, where $\varstepsize$ is computed as above and $M,\varconditionalexp$ are positive constants, which can be tuned to modify the decrease slope.

The third step size sequence, which we refer to as \emph{Conditional}, is defined in \ralg{conditional}. It consists of a \emph{Bootstrap phase} (up to iteration $\vartimeslot_{BS}$) in which step sizes remain constant, thus allowing broad exploration. Then an \emph{Adaptation} phase follows, up to iteration $M$, in which step sizes decrease, by default, linearly from an initial value $a$ to a final value $b$. This decrease is steeper than linear when the miss ratio measured at the current iteration is smaller than the $5$-th percentile of the miss ratio values observed so far. In this case the step size is halved, unless it already equals $b$. The intuition behind this phase is that we try to reduce the exploration extent every time we encounter a ``good'' allocation, i.e., an allocation that shows a small miss ratio compared to what we experienced so far. Note that we do not start immediately with the Adaptation phase, since we need to collect enough samples during the Bootstrap phase in order to correctly evaluate the quality of the current allocation. Finally, we continue with a \emph{Moderate} phase, in which step sizes are updated as above and are asymptotically vanishing, thus guaranteeing convergence.

After a preliminary evaluation, we set $\varconditionalexp=1/100$ as in~\cite{Wang2013f} and $b=a/10$. We set $\vartimeslot_{BS}$ and $M$, i.e., the duration of the bootstrap and adaptive phases, to the number of iterations in 6 minutes and 1 hour, respectively.  While the duration of these phases is clearly tied to the arrival rate, and are expected to require tuning when ported to a different scenario, we point out that performance achieved with these choices remains satisfying under the different scenarios we consider.

\subsection{Convergence}\label{res:conv}
\begin{figure}[t]
	\begin{centering}
	\includegraphics[width=0.22\textwidth, angle=270]{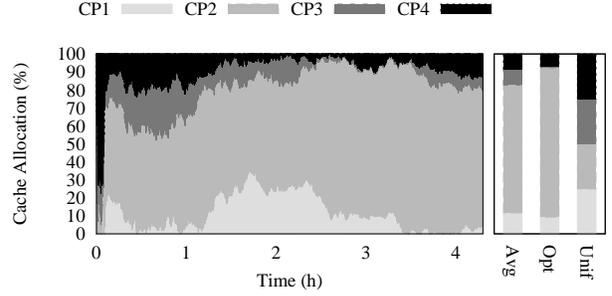}
	\end{centering}
	\caption{Evolution of the allocation of cache slots across CPs, with cache size $K=10^5$ and \emph{Conditional} step sizes. The three bars on the right represent the component-wise average of the allocated slots under \varalgoname allocation with Conditional steps (Avg),  the Optimal (Opt) and Uniform (Unif) allocations.}
	\label{fig:allocation_visual}
\end{figure}
We first consider a cache size of $K=10^5$ and 4 CPs, receiving 13\%, 75\%, 2\% and 10\% of requests, respectively. From \rfig{allocation_visual}, we can observe that, after a first exploration phase, the algorithm converges to a stable allocation. It is interesting to note that the average allocation (Avg), which is obtained by averaging each component of the allocation vector throughout the iterations, is very close to the optimal one, unlike the na\"ive uniform allocation policy.

Second, we consider a larger scenario with cache size $K=10^6$ and $p=10$ CPs, one of which we is a popular CP, to which 70\% of requests are directed, followed by a second one receiving 24\% of requests, other 6 CPs accounting for 1\% each and the remaining two CPs receiving no requests. 
\begin{figure}
	\begin{centering}
	\includegraphics[width=0.32	\textwidth, angle=270]{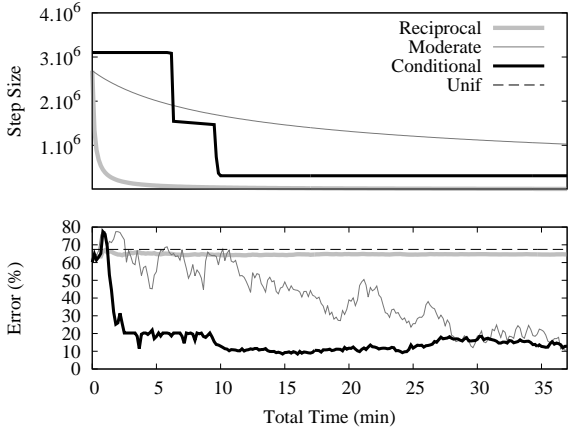}
	\end{centering}
	\caption{Error and step size sequence with cache size $K=10^6$.}
	\label{fig:convergence}
\end{figure}
\rfig{convergence} shows the step sizes and the inaccuracy of the algorithm, i.e. the distance to the optimal allocation, measured as:
\begin{equation}
	\label{eq:error}
	Error(\varallocvec) \defeq 
	\frac{\Vert \varallocvec^{OPT} - \varallocvec \Vert_\infty}{\varstorage} = 
	\frac{\max_{j=1\dots p} |\varallocation{\varCPindex} - \varallocation{\varCPindex}^{OPT}|}{\varstorage}.
\end{equation}

Observe that \emph{Reciprocal} steps decrease too fast, which immediately limits the adaptation of the allocation, significantly slowing down convergence. Conversely, \emph{Moderate} steps remain large for an overly long time, preventing the algorithm to keep the allocation in regions that guarantee good performance. \emph{Conditional} steps show the best performance since in the Adaptation phase the step sizes are sharply decreased if the current allocation is providing a small miss ratio.

\subsection{Sensitivity Analysis}\label{res:sens}

We next study how the performance of \varalgoname is affected by the algorithm parameters and the scenario.
We first focus on the time slot duration $\vartimeslotlength$. On the one hand, a small $\vartimeslotlength$ implies that only few requests are observed in each time slot, which may
result in a high noise $^{+}\varnoisevec\atstep{\vartimeslot},^{-}\varnoisevec\atstep{\vartimeslot}$, and ultimately affects the accuracy of the update. On the other hand,
a large $\vartimeslotlength$ decreases the measurement noise, but allows  updates to be made less frequently, which possibly slows down convergence. 

\begin{figure}[t]
	\begin{centering}
	\includegraphics[width=0.31\textwidth, angle=270]{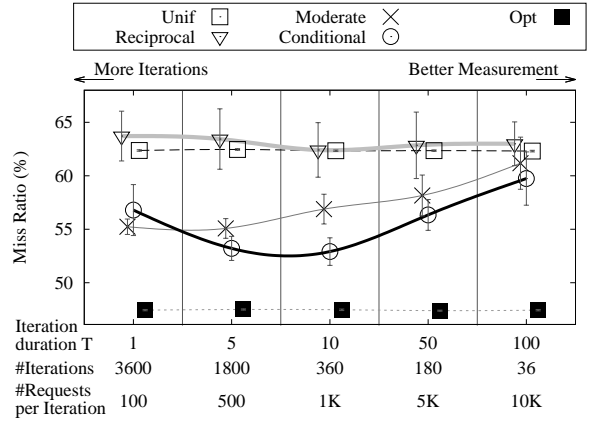}
	\end{centering}
	\caption{Impact of time-slot duration $\vartimeslotlength$ on the average miss ratio (bars represent the 95\% confidence interval over 20 runs).}
	\label{fig:misses_after_60_min}
\end{figure}

To evaluate the impact of $\vartimeslotlength$, \rfig{misses_after_60_min} shows the miss ratio measured over $1$h for the default scenario.
We consider \varalgoname with the three step size sequences, and compare it to the Uniform and to the Optimal allocations as benchmarks.
The figure shows that \varalgoname with the Conditional step size sequence enhances the cache efficiency  significantly.
We also observe that an iteration duration of $T=10s$  (corresponding to 100 samples on average per CP) represents a good compromise between
a more accurate miss ratio estimation based on more samples (with large $T$) and a larger number of iterations at the cost of lower accuracy (with small $T$).

\begin{figure}[t]
	\begin{centering}
	\includegraphics[width=0.45\columnwidth, angle=270]{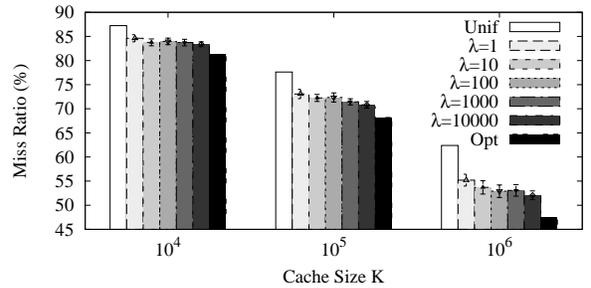}
	\end{centering}
	\caption{Miss rate measured over 1h for various average request rates $\lambda$ and cache sizes $\varstorage$.}
	\label{fig:misses}
\end{figure}

\rfig{misses} shows the cache miss rate measured over $1$h for a time slot length of $T=10$s and for various cache sizes $K\in\{10^4,10^5,10^6\}$ and arrival rates 
 $\lambda \in [1,10^4]$.
The figure confirms that the gains of \varalgoname hold for different cache sizes, and shows that the gain increases for large caches.
To interpret the results for different arrival rates, recall that for any given time slot duration $T$, the average request rate affects the measurement noise.
\rfig{misses} confirms that the  miss rate increases when the measurement noise is higher, i.e., for lower $\lambda$, but it also shows a very limited impact:
the number of time slots in a relatively short time (in $1$h, there are $360$ time slots of duration $T=10s$)
allows \varalgoname to converge to a good cache configuration, in spite of the noise and the consequent estimation errors.

\subsection{Changing Content Popularity}\label{res:nonstat}
\begin{figure}[t]
	\begin{centering}
	\includegraphics[width=0.32\textwidth, angle=270]{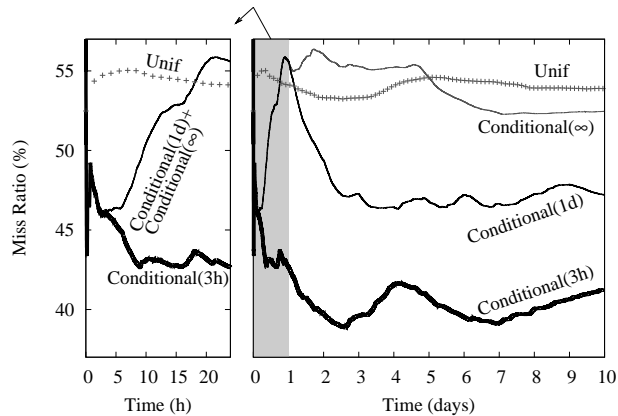}
	\end{centering}
	\caption{Miss ratio with varying content popularity. Step size sequences in \emph{Conditional($\tau$)}  are reset every $\tau$ interval. On the left the first day is zoomed, where \emph{Conditional($\infty$)} and \emph{Conditional($1d$)} correspond, since none of them have reinitialized the step sequence.}
	\label{fig:time_evolution}
\end{figure}

Recent studies~\cite{Garetto2015} have observed that the catalog statistics vary over time.
We show in this section that in order for \varalgoname to be robust to these variations, it suffices to periodically reinitialize the step sequence.
To model changing content popularity, we adopt the model of~\cite{Garetto2015}, in which each object is characterized by a sequence
of ON and OFF periods, with exponentially distributed duration $T_{ON}$ and $T_{OFF}$, respectively. At each time instant, an object can be ON or OFF, and only ON objects attract requests. As in~\cite{Garetto2015}, we set the catalog size to $3.5\cdot10^6$ and the cache size to $K=10^4$ objects.
We set the average ON and OFF duration to $1$ and $9$ days, respectively.
On average, we maintain the overall request rate of active objects equal to our default value $\lambda=100req/s$.

In \rfig{time_evolution} we compare \emph{Unif} and \emph{Conditional($\tau$)} 
that reinitialize the step sequence every $\tau$ amount of time.
We consider $\tau\in\{3h,1d,\infty\}$, i.e., 8\, reinitializations per day, daily reinitialization, or no-reinitialization, respectively.
As expected, reinitialization improves cache efficiency. Indeed, already after 3 hours of simulation, the evolution of the catalog misleads \emph{Conditional} and \emph{Conditional(1d)} (that overlap in this time interval) causing them to have performance worse than Unif. This is expected, since \emph{Conditional($\infty$)}  tries to converge to the optimal allocation, which is problematic in a non-stationary scenario. At the same time, it also shows that reinitializing step sequences as in \emph{Conditional($3h$)}
is sufficient to respond to the catalog dynamics.

\section{Conclusion}
\label{sec:conclusion}
One of the main challenges of in-network caching nowadays is its incompatibility with encrypted content.
Our work represents a first step in solving this challenge by proposing a simple and therefore appealing system design: \varalgonamelong\  requires solely the
knowledge of aggregated cache miss-intensities, based on which it provably converges to an allocation with a small optimality gap.
Simulation results show the benefits of the proposed algorithm under various scenarios, and  
 results obtained under complex content catalog dynamics further confirm the algorithm to be  applicable in scenarios of high practical relevance.


\section*{Acknowledgements}
This research was partially supported by Labex DigiCosme (ANR-11-LABEX-0045-DIGICOSME), part of the program ``Investissement d'Avenir'' Idex Paris-Saclay (ANR-11-IDEX-0003-02), and by the Swedish Foundation for Strategic Research through the Modane project. This work was carried out while Araldo was visiting KTH in the context of the EIT Digital project T15212 on ``Programmable data plane for Software Defined Networks''. This work benefited from support of NewNet@Paris, Cisco's Chair ``{\sc Networks for the Future}'' at Telecom ParisTech (\url{http://newnet.telecom-paristech.fr}). Any opinion, findings or recommendations expressed in this material are those of the author(s) and do not necessarily reflect the views of partners of the Chair.

\bibliographystyle{IEEEtran}
\bibliography{opencache}

\end{document}